\documentclass[12pt,british,epj]{webofc}
\usepackage[T1]{fontenc}
\usepackage[utf8]{inputenc}
\usepackage{amsthm}
\usepackage{amsmath}

\makeatletter
\theoremstyle{plain}
\newtheorem{thm}{\protect\theoremname}
  \theoremstyle{plain}
  \newtheorem{ax}[thm]{\protect\axiomname}

\usepackage[varg]{txfonts}
\wocname{EPJ Web of Conferences}
\woctitle{New Frontiers in Physics 2014}
\makeatother
\usepackage[matrix,frame,arrow]{xy}
\usepackage{amsmath}
\newcommand{\bra}[1]{\left\langle{#1}\right\vert}
\newcommand{\ket}[1]{\left\vert{#1}\right\rangle}
\newcommand{\qw}[1][-1]{\ar @{-} [0,#1]}

\newcommand{\gate}[1]{*{\xy *+<.6em>{#1};p\save+LU;+RU **\dir{-}\restore\save+RU;+RD **\dir{-}\restore\save+RD;+LD **\dir{-}\restore\POS+LD;+LU **\dir{-}\endxy} \qw}

\newcommand{\measureD}[1]{*{\xy*+=+<.5em>{\vphantom{\rule{0em}{.1em}#1}}*\cir{r_l};p\save*!R{#1} \restore\save+UC;+UC-<.5em,0em>*!R{\hphantom{#1}}+L **\dir{-} \restore\save+DC;+DC-<.5em,0em>*!R{\hphantom{#1}}+L **\dir{-} \restore\POS+UC-<.5em,0em>*!R{\hphantom{#1}}+L;+DC-<.5em,0em>*!R{\hphantom{#1}}+L **\dir{-} \endxy} \qw}

\newcommand{\multigate}[2]{*+<1em,.9em>{\hphantom{#2}} \qw \POS[0,0].[#1,0];p !C *{#2},p \save+LU;+RU **\dir{-}\restore\save+RU;+RD **\dir{-}\restore\save+RD;+LD **\dir{-}\restore\save+LD;+LU **\dir{-}\restore}
\newcommand{\ghost}[1]{*+<1em,.9em>{\hphantom{#1}} \qw}
\newcommand{\Qcircuit}[1][0em]{\xymatrix @*=<#1>} 

\newcommand{\prepareC}[1]{*{\xy*+=+<.5em>{\vphantom{#1\rule{0em}{.1em}}}*\cir{l^r};p\save*!L{#1} \restore\save+UC;+UC+<.5em,0em>*!L{\hphantom{#1}}+R **\dir{-} \restore\save+DC;+DC+<.5em,0em>*!L{\hphantom{#1}}+R **\dir{-} \restore\POS+UC+<.5em,0em>*!L{\hphantom{#1}}+R;+DC+<.5em,0em>*!L{\hphantom{#1}}+R **\dir{-} \endxy}}
\newcommand{\poloFantasmaCn}[1]{{{}^{#1}_{\phantom{#1}}}}

\makeatletter
\newcommand{\rA}{\mathrm{A}}
\newcommand{\rB}{\mathrm{B}}
\newcommand{\rC}{\mathrm{C}}
\newcommand{\rD}{\mathrm{D}}
\newcommand{\cC}{\mathcal{C}}
\newcommand{\cD}{\mathcal{D}}

\makeatother

\usepackage{babel}
  \providecommand{\axiomname}{Axiom}
\providecommand{\theoremname}{Theorem}

\begin{document}

\title{Conservation of information and the foundations of quantum mechanics}

\author{Giulio Chiribella \inst{1}\fnsep\thanks{\email{gchiribella@mail.tsinghua.edu.cn}}
\and Carlo Maria Scandolo \inst{1}\fnsep\thanks{\email{scandolocm10@mails.tsinghua.edu.cn}}\institute{Center for Quantum Information, Institute for Interdisciplinary Information Sciences, Tsinghua University, Beijing 100084, China}\abstract{We review a recent approach  to the foundations of quantum mechanics inspired  by  quantum information theory  \cite{Chiribella-purification,Chiribella-informational-derivation}.    The approach is based on a general framework, which  allows one to address a large class of physical theories which share  basic information-theoretic features.  We first illustrate  two very primitive features, expressed by   the axioms  of  causality and purity-preservation, which are  satisfied by both classical and quantum theory.             We then discuss  the  axiom of purification,  which expresses a strong version of the Conservation of Information and   captures the core of a vast number of protocols in quantum information.  Purification is a highly non-classical feature and leads directly to the emergence of entanglement at the purely conceptual level, without any reference to the superposition principle.        Supplemented by a few additional requirements, satisfied by classical and quantum theory, it provides   a complete  axiomatic characterization of quantum theory for finite dimensional systems.}}

\maketitle

\section{Introduction}

A new approach to the foundations of quantum theory has emerged over
the past three decades, drawing concepts and methods from the field
of quantum information \cite{fuchs,brassard}. This approach differs
from that of many previous works in quantum foundations, which were
primarily concerned with the interpretations of quantum mechanics
and with the measurement problem, maintaining the core of the Hilbert space
formalism untouched \cite{auletta}. Recent works try instead to \emph{derive}
the Hilbert space framework from more basic principles regarding information-processing.
In order to achieve this goal, one needs a more general framework
capable of describing possible alternatives to quantum mechanics.
These theories, called \emph{general probabilistic theories} \cite{Popescu-non-locality,Hardy-informational-1,Barrett,D'Ariano,Barnum-1,Chiribella-purification,Barnum-2,Chiribella-informational-derivation,hardy2011,Hardy-informational-2},
describe the experiments that can be performed with a given set of
physical devices, and provide a rule to assign probabilities to the
outcomes of such experiments. Compared to the tradition of quantum
logic \cite{mackey,ludwig,PironBook,foulisrandall}, which also aimed at
characterizing quantum theory into a larger landscape of theories,
the new approach differs in  the fact that it uses principles inspired
by information-theoretic protocols, such as quantum teleportation
\cite{Teleportation}. All the recent axiomatizations of quantum theory
\cite{Hardy-informational-1,D'Ariano,Goyal,Dakic,Masanes-1,Chiribella-informational-derivation,Hardy-informational-2,Masanes-2}
are clear examples of this new trend.

In this paper we provide a non-technical introduction to general probabilistic
theories, based on the framework established by D'Ariano, Perinotti,
and one of the authors in Refs.~\cite{Chiribella-purification,Chiribella-informational-derivation}.
This framework is particularly apt to capture the operational aspects
of a theory, making use of an intuitive graphical notation borrowed
from the area of categorical quantum mechanics \cite{cqm,Coecke-Kindergarten,Coecke-Picturalism,Selinger}.
After introducing the framework, we discuss the axiomatization of
quantum theory presented in Ref.~\cite{Chiribella-informational-derivation}.
In particular, we focus on three axioms, which we consider particularly fundamental.
The first two axioms are Causality and Purity-Preservation, which
are satisfied by both classical and quantum theory. The third  axiom is Purification, which is not satisfied by classical theory
and is responsible for many of the surprising features of quantum
information. In particular, we show that purification leads directly
to the no-cloning theorem \cite{noclon} and to the phenomenon of
entanglement.

\section{Why probabilistic theories}

Before entering into details, let us have a brief discussion on what
the framework of general probabilistic theories aims to accomplish.
After all, why going through the trouble of exploring more general
theories, when quantum mechanics is already so successful in its predictions?
In short, one can identify four reasons:
\begin{enumerate}
\item \emph{Contribution to a deeper understanding of quantum mechanics.}\\
Reconstructing a theory from basic physical principles, rather having
just a mathematical description, helps build intuition and promote
the advancement of the theory itself. Think for example of Einstein's
reconstruction of Lorentz transformations from the principle of relativity
and from the law of light propagation \cite{rindler}. 
\item \emph{Extensions and modifications of quantum mechanics.}\\
Despite the present success of quantum mechanics, it is conceivable
that the theory may need modifications in new regimes that have not
been explored yet. The analysis of more general theories helps suggest
which quantum mechanical axioms can be modified to adapt the theory
to new scenarios, such as those of a perspective theory of quantum
gravity. 
\item \emph{Search for links between quantum information protocols.}\\
Quantum information theorists have devised a multitude of new protocols,
which turn the counter-intuitive features of quantum mechanics into
advantages \cite{Nielsen-Chuang,Wilde}. A natural tendency is then
to try to recognize the underlying patterns and to establish direct
links between different quantum protocols. Besides the benefit of
conceptual clarification, this may also help devise new protocols. 
\item \emph{Effective restrictions of quantum mechanics.}\\
Suppose not all quantum states allowed by quantum mechanics are accessible
with a particular experimental setup. For example, linear optics techniques
can easily generate and manipulate Gaussian states of light, but are
not able to access non-Gaussian states and operations. Given an effective
theory describing a restricted subset of quantum states and operations,
a natural question is: ``What quantum features can be observed?''.
One way to provide an answer is to phrase the theory as a general
probabilistic theory and check which axioms are satisfied.
\end{enumerate}
For a more extended presentation and for more arguments we refer the
reader to the insightful discussion by Hardy and Spekkens \cite{Hardy-Spekkens}.

\section{Systems and tests}

We now discuss a framework for general probabilistic theories, following
the scheme of Ref.~\cite{Chiribella-purification,Chiribella-informational-derivation}
(see also \cite{Chiribella-divulgative}). In this framework there
are two primitive notions, the notion of \emph{physical system} and
the notion of \emph{test}.

A \emph{test} represents a use of a physical device (e.g.\,a beam-splitter,
a polarimeter, or a Stern-Gerlach apparatus). Every device has an
input system and an output system. We denote systems with capital
letters, such as $\mathrm{A}$, $\mathrm{B}$, and so on. Among all
systems, it is convenient to include the \emph{trivial system}, which
represents ``nothing''. A device with trivial system as input is
a device with no input, and a device with trivial system as output
is a device with no output.

In general, a device can have various outcomes, which can be e.g.\,a
sequence of digits, or a spot in a photographic plate. The outcomes
can be identified by the experimenter, and each outcome corresponds
to a different process that can take place when the device is used.
Hence, tests will be represented as collections of processes labelled
by outcomes, such as $\left\{ \mathcal{C}_{i}\right\} $. We will
also adopt a graphical language, which is intuitive and at the same
time mathematically rigorous \cite{cqm,Coecke-Kindergarten,Coecke-Picturalism,Selinger}.
Using this language, a test is depicted as a box with incoming and
outgoing wires that represent the input and output system respectively.
For example, the test $\left\{ \mathcal{C}_{i}\right\} $ will be
represented as\[
\begin{aligned}\Qcircuit @C=1em @R=.7em @!R {    & \qw \poloFantasmaCn{\rA} &  \gate{\{\cC_{i}\}} & \qw \poloFantasmaCn{\rB} &\qw}\end{aligned}~.
\]If we want to specify which process occurred, we omit the braces,
as in the following\[
\begin{aligned}\Qcircuit @C=1em @R=.7em @!R {    & \qw \poloFantasmaCn{\rA} &  \gate{\cC_{i}} & \qw \poloFantasmaCn{\rB} &\qw}\end{aligned}~.
\]Deterministic evolutions, such as the unitary evolution induced by
Schrödinger equation, can be represented in this framework as tests
with only one possible outcome (where the meaning of the outcome is
just that the evolution took place).

The process of preparing a state can also be described as a test,
specifically a test with the trivial system as input and the system
that is being prepared as output. A test of this form, say $\left\{ \rho_{i}\right\} $,
is called \emph{preparation-test} and represents a device which prepares
the system in a state $\rho_{i}$, randomly chosen from the set $\left\{ \rho_{i}\right\} $.
We represent a preparation-test as\[
\begin{aligned}\Qcircuit @C=1em @R=.7em @!R {    &  \prepareC{\{\rho_{i}\}} & \qw \poloFantasmaCn{\rA} &\qw}\end{aligned}~.
\]  On the other hand, also destructive measurements can be represented
as tests. These special tests have no output (trivial system as output),
and destroy the input system while acquiring some information from
it, as it happens e.g.\,when an electron is absorbed by a photographic
plate, leaving a spot on it. A test of this form, say $\left\{ a_{i}\right\} $,
is called \emph{observation-test} and each individual process $a_{i}$
represents a way of destroying the input system. Using graphical language,
we represent an observation-test as\[
\begin{aligned}\Qcircuit @C=1em @R=.7em @!R {   & \qw \poloFantasmaCn{\rA} &\measureD{\{a_{i}\}}}\end{aligned}~.
\]

\section{\label{sec:Sequential-and-parallel}Sequential and parallel composition}

Having fixed how to represent devices, the next step is to describe
how to connect them. Devices can be connected in sequence or in parallel.
In sequential composition the two devices are connected one after
the other. To do so, clearly the input of the second device must be
the same as the output of the first device, as shown in the following
example:\begin{equation}\label{aaa}
\begin{aligned} \Qcircuit @C=1em @R=.7em @!R {  \prepareC{\{\rho_{i}\}}  & \qw \poloFantasmaCn{\rA} &  \gate{\{\cC_{j}\}} & \qw \poloFantasmaCn{\rB} &\measureD{\{b_{k}\}}}  \end{aligned}~.
\end{equation}The above diagram gives instructions on how to build up an experiment:
in this experiment one first initializes system $\mathrm{A}$ with
the preparation-test $\left\{ \rho_{i}\right\} $, then performs the
test $\left\{ \mathcal{C}_{j}\right\} $, which transforms system
$\mathrm{A}$ into system $\mathrm{B}$, and finally one acquires
information from $\mathrm{B}$ by performing the observation-test
$\left\{ b_{k}\right\} $. If we wish to express which events actually
occurred, we write\begin{equation} \label{probability}
\begin{aligned}\Qcircuit @C=1em @R=.7em @!R {  \prepareC{\rho_{i}}  & \qw \poloFantasmaCn{\rA} &  \gate{\cC_{j}} & \qw \poloFantasmaCn{\rB} &\measureD{b_{k}}}\end{aligned}~.
\end{equation}This means that the state $\rho_{i}$ was prepared, the process $\mathcal{C}_{j}$
took place, and finally the system was destroyed, producing the outcome
$k$.

We denote the sequential composition of a process $\mathcal{D}_{j}$
after a process $\mathcal{C}_{i}$ as $\mathcal{D}_{j}\circ\mathcal{C}_{i}$.
In general, note that there is a strict ordering in sequential composition:
some tests are performed first and other later. In the graphical language,
the ordering goes from left to right. This ordering will be essential
to phrase the causality axiom (section~\ref{sec:Causality}), which
forbids signalling from the future to the past.

Let us move now to parallel composition. Parallel composition of tests
arises when we apply two devices to two different systems independently.
The parallel composition of two processes is denoted by $\mathcal{C}_{i}\otimes\mathcal{D}_{j}$ and simply represented as
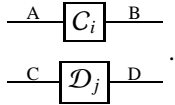
\[
\begin{aligned}\Qcircuit @C=1em @R=.7em @!R {     & \qw \poloFantasmaCn{\rA}  & \gate{\cC_{i}} & \qw \poloFantasmaCn{\rB} &\qw  \\    & \qw \poloFantasmaCn{\rC}  &\gate{\cD_{j}} & \qw \poloFantasmaCn{\rD} &  \qw }\end{aligned}~.
\]  
An important difference between sequential and parallel composition
is that, when two processes are composed in parallel, the order in
which they take place does not matter. 

A particular case of parallel composition is the composition of preparation
devices. When a preparation device prepares system $\mathrm{A}$ in
a state $\rho$ and another device prepares system $\mathrm{B}$ in
a state $\sigma$, we say that the composite system $\mathrm{AB}$
is in a \emph{product state}, denoted by $\rho\otimes\sigma$ and
graphically represented as
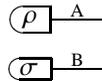
\[
\begin{aligned}\Qcircuit @C=1em @R=.7em @!R { & \prepareC{\rho}    & \qw \poloFantasmaCn{\rA} &  \qw   \\  & \prepareC{\sigma}    & \qw \poloFantasmaCn{\rB}  &  \qw }\end{aligned}~.
\] It is important to note that the operations that can be performed
on a composite system are not restricted to product operations. In
general, one can also use a joint device which processes the component
systems together. Such a process represents the result of an \emph{interaction},
such as e.g.\,the interaction between two beams of particles in an
accelerator. Joint devices will be represented as boxes with multiple
wires, one wire for each system, as in the following example:
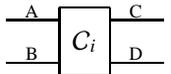
\[
\begin{aligned}\Qcircuit @C=1em @R=.7em @!R {     & \qw \poloFantasmaCn{\rA}  & \multigate{1}{\cC_{i}} & \qw \poloFantasmaCn{\rC} &\qw  \\    & \qw \poloFantasmaCn{\rB}  &\ghost{\cC_{i}} & \qw \poloFantasmaCn{\rD} &  \qw }\end{aligned}~.
\]

\section{A consistent rule to predict probabilities}

When we have a diagram with no external wires, like diagram~\eqref{probability},
we interpret it as a \emph{probability}. This is a shorthand notation
to mean that a process that starts with the preparation of a state
and ends with the destruction of the system yields a probability.
For example, diagram~\eqref{probability} represents the \emph{joint
probability} that the state $\rho_{i}$ is prepared, the transformation
$\mathcal{C}_{j}$ takes place, and the process $b_{k}$ destroys
the system.

The rule to compute the probabilities of all possible diagrams with
no external wires is assumed as part of the specification of the theory.
The only requirements for this rule are
\begin{enumerate}
\item the sum of the probabilities for all the outcomes produced in an experiment
must be equal to 1;
\item the outcome probabilities for experiments performed in parallel must
be of the product form.
\end{enumerate}

\section{Purity of states and transformations}

In both classical and quantum statistical mechanics it is common to
distinguish between pure and mixed states. For example, in the quantum
case, pure states are described by rays in the Hilbert space of the
system, while mixed states are described by density operators. The
distinction between pure and mixed states, however, is not specific
to quantum or classical theory: in fact, it makes sense in every probabilistic
theory. More generally, one can define also pure and mixed transformations. 

The idea at the basis of the definition is \emph{coarse-graining}.
Let us clarify it with an easy example. In the roll of a die, there
are six basic outcomes, identified by the numbers from 1 to 6. However,
we can consider a coarse-grained outcome, which results from joining
together some of the basic outcomes, neglecting some information.
This is the case, for example, when we just say that the outcome of
the roll was an odd number. This coarse-grained outcome is the union
of the basic outcomes 1, 3 and 5.

Clearly, after doing a coarse-graining we lose some information. We
can say that a transformation is \emph{pure} if it does not arise
as a coarse-graining of other transformations; in this way, a pure
transformation represents a process on which we have maximal information.
An example of pure transformation in quantum theory is the unitary
evolution resulting from Schrödinger equation. The definition of pure
transformation also applies to states, which are a particular type
of transformations, namely the transformations implemented by preparation
devices. A state that is not pure is called \emph{mixed}: when a system
is described by a mixed state, one has only partial information about
the preparation. For instance, we may know that some pure states $\psi_{i}$'s
are prepared with given probabilities $p_{i}$'s. If we ignore which
state $\psi_{i}$ has been prepared, we describe the system with the
mixed state $\rho=\sum_{i}p_{i}\psi_{i}$, which can be regarded as
a sort of ``expectation state'' of the system. Here we are using
the symbol of sum just as a notation for probabilistic coarse-graining.
However, with a little work, one can actually define a suitable notion
of sum of transformations, using the rule for probabilities that is
provided by the theory \cite{Chiribella-purification}.

\section{Purity-preservation}

Equipped with the notions of pure state and pure transformation, we
can now discuss one of the axioms of Ref.~\cite{Chiribella-informational-derivation}.
This axiom provides an answer to the following question: ``Is the
composition of two pure transformations still pure?''. Intuitively,
when we have maximal knowledge of two processes, we should also have
maximal knowledge of the process that results from their composition.
This intuitive requirement is formalized by the axiom of \emph{purity-preservation}.
\begin{ax}[Purity-preservation]
The sequential and parallel composition of pure transformations yields
pure transformations.
\end{ax}
Purity-preservation is a very primitive requirement. Think of the
theory as an algorithm, used by a physicist to make deductions from
known facts: given as datum that system $\mathrm{A}$ undergoes the
process $\mathcal{C}$ from time $t_{0}$ to $t_{1}$ and the process
$\mathcal{D}$ from time $t_{1}$ to $t_{2}$, the algorithm deduces
that system $\mathrm{A}$ undergoes the process $\mathcal{D}\circ\mathcal{C}$
from time $t_{0}$ to $t_{2}$. The lack of purity-preservation would
mean that the algorithm is not able to determine what really happened
to the system after a sequence of time-steps, even when provided with
the most precise input about each individual step. Not even quantum
mechanics is so random: for example, the composition of two unitary
evolutions is still a unitary evolution, and not a stochastic process
with multiple outcomes. More generally, it is easy to see that both
classical and quantum theory satisfy the axiom of purity-preservation.

\section{\label{sec:Causality}Causality}

Another very primitive requirement about physical theories is causality.
One can phrase the axiom as follows.
\begin{ax}[Causality]
The probability of an outcome at a certain step does not depend on
the choice of experiments performed at later steps.
\end{ax}
The ``steps'' mentioned here are the steps in a sequence of laboratory
operations, such as those depicted in example~\eqref{aaa}. In that
particular example, the causality axiom ensures that the probability
that system $\mathrm{A}$ is prepared in the state $\rho_{i}$ does
not depend on the choice of test $\left\{ \mathcal{C}_{j}\right\} $
or on the choice of the destructive measurement $\left\{ b_{k}\right\} $.
Informally, the causality axiom states that it is impossible to signal
from the future to the past. It is easy to see that both classical
theory and quantum theory fulfil this requirement.

The impossibility of signalling from the future to the past implies
the impossibility of instantaneous signalling across space \cite{Chiribella-purification}.
Suppose that two distant parties, conventionally called Alice and
Bob, perform two independent tests $\left\{ \mathcal{A}_{i}\right\} $
and $\left\{ \mathcal{B}_{j}\right\} $ on their respective systems
in their laboratories. Since the two tests are performed in parallel,
the order does not matter, i.e.\,the probability of the outcomes
does not depend on whether Alice or Bob performs her/his test first.
Combining this observation with the causality axiom, we have that
the probability that Alice finds outcome $i$ must not depend on the
choice of test $\left\{ \mathcal{B}_{j}\right\} $ performed by Bob,
and vice versa. Of course, there can be correlations between Alice's
and Bob's outcomes, if the state $\rho_{\mathrm{AB}}$ of the composite
system is not of the product form $\rho_{\mathrm{A}}\otimes\rho_{\mathrm{B}}$.
These correlations will be described by some joint probability distribution,
$p_{\mathrm{AB}}\left(i,j\right)$, but Alice's marginal probability
distribution $p_{\mathrm{A}}\left(i\right)=\sum_{j}p_{\mathrm{AB}}\left(i,j\right)$
is independent of the choice of test $\left\{ \mathcal{B}_{j}\right\} $,
and Bob's marginal probability distribution $p_{\mathrm{B}}\left(j\right)=\sum_{i}p_{\mathrm{AB}}\left(i,j\right)$
is independent of the choice of test $\left\{ \mathcal{A}_{i}\right\} $.
In other words, the correlations that arise from a causal theory do
not allow for signalling across space. As a consequence, if Alice
wants to send a message to Bob, she has to send some physical system.
This also implies that the speed of every message is limited by the
maximum speed at which physical systems can be transferred in space.
For example, if we assume that the maximum speed coincides with the
speed of light in vacuum, we have that faster-than-light communication
is ruled out in every probabilistic theory that satisfies the causality
axiom.

\section{The purification principle and the conservation of information}

So far we have discussed axioms that are satisfied by both classical
and quantum theory. However, in order to identify quantum theory one
needs at least one axiom that is not satisfied by classical theory.
Such an axiom should capture what makes the quantum world so radically
different from the classical one, and possibly, should allow one to
deduce the key protocols of quantum information theory. One axiom
that possesses these features is the purification axiom \cite{Chiribella-purification,Chiribella-informational-derivation}.
\begin{ax}[Purification]
Every physical process can be simulated in an essentially unique
way as a reversible evolution of the system interacting with a pure
environment.
\end{ax}
Let us unpack the content of this statement. First, the axiom tells
us that every process, even an irreversible one, can be modelled as
the result of a reversible process, where the system interacts with
the environment. The origin of irreversibility is only in the fact
that the environment is discarded: at least in principle, if the experimenter
were able to maintain full control of the degrees of freedom that
are interacting during the experiment, the overall evolution would
be reversible. This fact expresses the principle of Conservation of
Information, which states that, at some fundamental level, information
cannot be destroyed, but can only be discarded. The Conservation of
Information, \emph{per se}, is not a distinctive quantum feature:
for example, it is satisfied also by Newton's equations of motion
and by other dynamical equations in classical physics. What is distinctive
about quantum theory is the combination of the Conservation of Information
with the ``pure environment'' part of the purification axiom: when
we model an irreversible process as a reversible evolution of the
system along with the environment, we can always start with the environment
in a pure state. In classical physics, it is possible to simulate
a stochastic process through a reversible process, but the price for
such a simulation is that one needs an external source of randomness,
provided by the environment. Instead, the purification axiom imposes
that even the stochastic processes can be simulated without the need
of initial randomness. Consider, for example, the process of preparing
a system $\mathrm{A}$ in a mixed state $\rho$. In this case, the
purification axiom ensures that we can prepare $\rho$ with the following
procedure:
\begin{enumerate}
\item prepare $\mathrm{A}$ in a pure state $\alpha$ and prepare another
system $\mathrm{E}$ in a pure state $\eta$;
\item let the composite system $\mathrm{AE}$ evolve with a suitable reversible
process $\mathcal{U}$, thus obtaining the pure state $\Psi_{\mathrm{AE}}=\mathcal{U}\left(\alpha\otimes\eta\right)$;
\item discard system $\mathrm{E}$.
\end{enumerate}
Here system $\mathrm{E}$ plays the role of the environment and the
pure state $\Psi$ is called a \emph{purification} of $\rho$. For
example, in quantum theory it is possible to prepare a 2-level system
in the mixed state $\rho=\frac{1}{2}\left(\ket{0}\bra{0}+\ket{1}\bra{1}\right)$
by first preparing two systems in the pure product state $\ket{0}\ket{0}$
and then letting them evolve with a suitable unitary evolution that
transforms the state $\ket{0}\ket{0}$ into the singlet state $\ket{S}=\frac{1}{\sqrt{2}}\left(\ket{0}\ket{1}-\ket{1}\ket{0}\right)$.
By discarding the second system, one remains with the first system
in the mixed state $\rho=\frac{1}{2}\left(\ket{0}\bra{0}+\ket{1}\bra{1}\right)$.
By contrast, in classical theory it is impossible to simulate the
preparation of a mixed state using only pure states and reversible
evolutions.

More generally, the purification axiom guarantees that every physical
process $\mathcal{C}$, transforming the state of system $\mathrm{A}$
from $\rho_{\mathrm{A}}$ to $\rho'_{\mathrm{A}}=\mathcal{C}\left(\rho_{\mathrm{A}}\right)$,
can be simulated according to the same procedure described for the
preparation of mixed states.

The third and last important point about the purification axiom is
that the reversible simulation of a physical process is ``essentially
unique'': if two reversible evolutions on system $\mathrm{AE}$,
say $\mathcal{U}$ and $\mathcal{U}'$, simulate the same process,
then there exists a reversible evolution $\mathcal{V}_{\mathrm{E}}$,
acting only on the environment, such that $\mathcal{U}'=\left(\mathcal{I}_{\mathrm{A}}\otimes\mathcal{V}_{\mathrm{E}}\right)\circ\mathcal{U}$,
where $\mathcal{I}_{\mathrm{A}}$ is the identity on system $\mathrm{A}$.
In other words, this means that all the reversible evolutions that
simulate the same process on system $\mathrm{A}$ must be equivalent,
up to a ``gauge transformation'' on the environment. In quantum
theory, this means that two unitary operators $U$ and $U'$, acting
on the Hilbert space $\mathcal{H}_{\mathrm{A}}\otimes\mathcal{H}_{\mathrm{E}}$
and simulating the same process on $\mathrm{A}$, must be equal up
to a change of basis in the Hilbert space $\mathcal{H}_{\mathrm{E}}$.
The uniqueness up to reversible transformations is a very important
feature, because it guarantees that all the models that we can invent
to account for the irreversibility of a process are physically equivalent.

\section{The no-cloning theorem}

Among the appealing features of the purification axiom there is the
fact that it gives direct access to many of the key structures of
quantum information \cite{Chiribella-purification}. To give the flavour
of how this is accomplished, here we show how the purification axiom
can be used to derive the no-cloning theorem \cite{noclon} in the
context of general probabilistic theories. 

Suppose we want to construct a copy machine, which takes a system
$\mathrm{A}_{1}$ in some pure state $\alpha$ and another identical
system $\mathrm{A}_{2}$ in a fixed state $\alpha_{0}$ as input,
providing systems $\mathrm{A}_{1}$ and $\mathrm{A}_{2}$ in the state
$\alpha\otimes\alpha$ as output. If the machine works for every possible
pure state $\alpha$, then we call it a \emph{universal copy machine}.
While in classical physics there is no limitation in principle about
the construction of universal copy machines, in quantum physics one
has the no-cloning theorem \cite{noclon}, stating that no physical process can make
two perfect replicas of an arbitrary pure state.

Let us see how this result can be directly derived from our axioms.
The proof is by contradiction: suppose there exists a process that
transforms $\alpha\otimes\alpha_{0}$ into $\alpha\otimes\alpha$
for every pure state $\alpha$. By the purification axiom, this process
can be realized by combining the input systems with another system
$\mathrm{E}$ (the environment) in some pure state, and letting them
reversibly evolve via some process $\mathcal{U}$, thus obtaining
the state $\mathcal{U}\left(\alpha\otimes\alpha_{0}\otimes\eta\right)$.
Now, by definition of copy machine, after discarding system $\mathrm{E}$,
systems $\mathrm{A}_{1}$ and $\mathrm{A}_{2}$ should be in the state
$\alpha\otimes\alpha$. This in particular implies that system $\mathrm{A}_{1}$
must be in the state $\alpha$. Hence, from the input to the output,
the state of system $\mathrm{A}_{1}$ undergoes the identity transformation
$\alpha\mapsto\alpha$  \footnote{Here we make the mild assumption that the identity transformation is the \emph{only} physical transformation that maps every pure state into itself.  This fact can be reduced to other principles, such as local tomography or a weaker property known as \emph{local tomography on pure states} \cite{Chiribella-purification}.}. Summarizing, the identity transformation on
$\mathrm{A}_{1}$ can be realized by 
\begin{enumerate}
\item combining system $\mathrm{A}_{1}$ with system $\mathrm{A}_{2}\mathrm{E}$
in the state $\alpha_{0}\otimes\eta$;
\item applying the reversible evolution $\mathcal{U}$ on $\mathrm{A}_{1}\mathrm{A}_{2}\mathrm{E}$;
\item discarding system $\mathrm{A}_{2}\mathrm{E}$.
\end{enumerate}
On the other hand, another (trivial) way to realize the identity transformation
on $\mathrm{A}_{1}$ is to combine it with the system $\mathrm{A}_{2}\mathrm{E}$
in the state $\alpha_{0}\otimes\eta$, apply the reversible transformation
$\mathcal{U}'=\mathcal{I}_{\mathrm{A}_{1}}\otimes\mathcal{I}_{\mathrm{A}_{2}}\otimes\mathcal{I}_{\mathrm{E}}$,
and discard system $\mathrm{A}_{2}\mathrm{E}$. Since the reversible
simulation of physical processes is unique up to reversible transformations
on the environment, one must have $\mathcal{U}'=\left(\mathcal{I}_{\mathrm{A}_{1}}\otimes\mathcal{V}_{\mathrm{A}_{2}\mathrm{E}}\right)\circ\mathcal{U}$
for some reversible transformation $\mathcal{V}_{\mathrm{A}_{2}\mathrm{E}}$
acting only on $\mathrm{A}_{2}\mathrm{E}$. Equivalently, this means
that $\mathcal{U}=\left(\mathcal{I}_{\mathrm{A}_{1}}\otimes\mathcal{V}_{\mathrm{A}_{2}\mathrm{E}}^{-1}\right)\circ\mathcal{U}'\equiv\mathcal{I}_{\mathrm{A}_{1}}\otimes\mathcal{V}_{\mathrm{A}_{2}\mathrm{E}}^{-1}$,
since $\mathcal{U}'$ is the identity on system $\mathrm{A}_{1}\mathrm{A}_{2}\mathrm{E}$.
Using this relation, we obtain
\[
\mathcal{U}\left(\alpha\otimes\alpha_{0}\otimes\eta\right)=\alpha\otimes\Psi_{\mathrm{A}_{2}\mathrm{E}},\qquad\qquad\Psi_{\mathrm{A}_{2}\mathrm{E}}:=\mathcal{V}_{\mathrm{A}_{2}\mathrm{E}}^{-1}\left(\alpha_{0}\otimes\eta\right).
\]
Note that, by definition, the (pure) state $\Psi_{\mathrm{A}_{2}\mathrm{E}}$
is independent of $\alpha$. Hence, we reached a contradiction: if
we discard systems $\mathrm{A}_{1}$ and $\mathrm{E}$ on both sides,
the l.h.s. is equal to $\alpha$ (by the hypothesis that $\mathcal{U}$
realizes a copying process) and the r.h.s. is independent of $\alpha$.
In conclusion, we proved that the process $\alpha\otimes\alpha_{0}\mapsto\alpha\otimes\alpha$
cannot be realized in any theory satisfying the purification principle. 

Conceptually, proving the no-cloning theorem directly from purification
is an important result: it tells us that the existence and essential
uniqueness of a reversible simulation of physical processes implies
the impossibility of universal copying machines. It is also important
to recall that the no-cloning theorem is the working principle at
the basis of the security of quantum cryptographic protocols, such
as the BB84 key distribution protocol \cite{QKD}. Having derived
this theorem from first principles suggests that one may be able to
provide also an axiomatic proof of the security of key distribution
based on the purification axiom.

\section{Entanglement\label{sec:Entanglement}}

Entanglement is one of the weirdest features of quantum mechanics
\cite{EPR,Schroedinger}. It gives rise to correlations that cannot
be explained by any local realistic model \cite{Bell} and deeply
challenge our intuition about the microscopic world. However, besides
being puzzling, entanglement is also a precious resource for quantum
communication protocols \cite{Nielsen-Chuang,Wilde}, such as quantum
teleportation \cite{Teleportation}.

In quantum mechanics, entanglement appears as a mathematical consequence
of the superposition principle. But is there a deeper reason for its
existence? In order to answer this question, we need first to provide
a theory-independent definition of entanglement. In section~\ref{sec:Sequential-and-parallel},
we introduced product states as the result of independent preparation
operations performed in parallel for different systems. When two systems
$\mathrm{A}$ and $\mathrm{B}$ are in a pure product state, say $\alpha\otimes\beta$,
one can assign a pure state to each system: system $\mathrm{A}$ is
in the state $\alpha$ and system $\mathrm{B}$ is in the state $\beta$.
By contrast, we say that a pure state $\Psi$ is \emph{entangled}
if it is \emph{not} a product state. When systems $\mathrm{A}$ and
$\mathrm{B}$ are in an entangled state, one has maximal knowledge
about the composite system $\mathrm{AB}$, without having maximal
knowledge of its parts. The above definition of entanglement captures
an idea expressed by Schrödinger \cite{Schroedinger}, who famously
wrote
\begin{quotation}
``the best possible knowledge of a whole does \emph{not} \emph{necessarily}
include the best possible knowledge of all its parts''.
\end{quotation}
It is easy to see that every probabilistic theory satisfying the purification
axiom must have entangled states \cite{Chiribella-purification}.
Indeed, suppose the pure states of system $\mathrm{AB}$ are only
of the product form $\alpha\otimes\beta$ for some pure states $\alpha$
and $\beta$ of systems $\mathrm{A}$ and $\mathrm{B}$ respectively.
Then, when we discard system $\mathrm{B}$, the remaining state of
system $\mathrm{A}$ is pure. In conclusion, the composite system
$\mathrm{AB}$ cannot be used to purify any mixed state of $\mathrm{A}$.
In order for the purification axiom to hold, there must exist at least
a system $\mathrm{B}$ such that the composite system $\mathrm{AB}$
is in an entangled state. In summary, the purification axiom leads
directly to entanglement. Conceptually, this is also an important
point, because entanglement gives rise to the most dramatic differences
between quantum and classical theory. In the same paper quoted above,
Schrödinger expressed the intuition that entanglement is at the centre
of the structure that characterizes quantum mechanics, writing
\begin{quotation}
``I would not call that \emph{one} but rather \emph{the} characteristic
trait of quantum mechanics, the one that enforces its entire departure
from classical lines of thought''.
\end{quotation}
In a sense, the axiomatization of finite-dimensional quantum theory
presented in Ref.~\cite{Chiribella-informational-derivation} fulfills
this intuition: the basic rules of the Hilbert space framework can
be derived from the purification axiom (which arguably captures the
structures that Schrödinger was highlighting in his paper), in combination
with five other axioms that are satisfied also by classical theory,
such as e.g.\,the axioms of causality and purity-preservation.

\section{Conclusions}

In this paper we reviewed a basic language for general probabilistic
theories and illustrated three operational axioms that can be expressed
in this language, namely purity-preservation, causality, and purification.
In a nutshell: purity-preservation is the requirement that maximal
knowledge of the processes happening in a sequence of time-steps implies
maximal knowledge of the overall process from the input to the output.
Causality is the requirement that no signal can be sent from the future
to the past. Purification is the requirement that every physical process
can be simulated in an essentially unique way as a reversible process
where the system interacts with an environment, initially prepared
in a pure state. Purification expresses a strengthened form of the
principle of Conservation of Information: it guarantees that one can
always account for irreversibility by formulating a model where, at
the fundamental level, information is preserved. Moreover, it guarantees
that such a model is essentially unique and does not require a source
of randomness in the environment. Purification leads directly the
key structures in quantum mechanics, like the no-cloning theorem and
the existence of entangled states. Combined with purity-preservation,
causality, and other three axioms satisfied also by classical theory,
purification leads to a complete axiomatization of quantum theory
in finite dimension. The main message emerging from this derivation
is that the core of quantum theory can be identified in the ability
to simulate irreversible and stochastic processes using only pure
resources and reversible evolutions.

\bigskip{}

\textbf{Acknowledgements.} This work was supported by the National
Basic Research Program of China (973) 2011CBA00300 (2011CBA00301),
by the National Natural Science Foundation of China (Grants 11350110207,
61033001, 61061130540), by the 1000 Youth Fellowship Program of China.
GC acknowledges the hospitality of Perimeter Institute for Theoretical
Physics. Research at Perimeter Institute for Theoretical Physics is
supported in part by the Government of Canada through NSERC and by
the Province of Ontario through MRI.

\bibliographystyle{plain}

\end{document}